\documentclass[twoside]{article}
\usepackage{latexsym,amsmath,amssymb}
\usepackage{hyperref}
\usepackage{cite} 
\usepackage{sectsty}
  \sectionfont{\large} \subsectionfont{\normalsize}
\usepackage{graphicx}

\usepackage{theorem}
\usepackage[all]{xy}

{\theorembodyfont\rmfamily

}

\date{}

\title{\vspace{-9ex}
{\centering
 \textbf{\large Analytical Pricing of 2 Factor Structural PDE model for a Puttable Bond with Credit Risk
}}}

 \vspace{-4ex}

\author{\small\textsf{\bfseries 
$^{1}$~ Hyong-Chol O, $^{2}$~Dae-Song Choe, $^{3}$~Gyong-Dok Rim}\\[-.5ex]
{\footnotesize  ${}^{1, 2, 3}$ Faculty of Mathematics, \textbf{Kim Il Sung} University,}
{\footnotesize   Pyongyang , D P R K}\\[-.5ex]
{\footnotesize * Corresponding author's Email address: $^{1}$hc.o@ryongnamsan.edu.kp }}

\begin{document}

\maketitle
\thispagestyle{empty}

\vspace{-.6cm}

\noindent \textbf{Abstract: }In this paper is proposed a 2 factor structural PDE model of pricing puttable bond with credit risk and derived the analytical pricing formula. To this end, first, a 2 factor structural (PDE) model of pricing zero coupon bond with credit risk is provided, the analytical pricing formula is derived under some conditions for default boundary and default recovery, and the strict monotonicity of the bond price function with respect to the firm value variable is proved. Then a (2 factor) pricing model of the option on zero coupon bond with credit risk is provided and under some condition on the exercise price its analytical pricing formula is derived by transforming the 2 factor model into a terminal boundary value problem for Black-Scholes equation with time dependent coefficient using zero coupon bond as numeraire. Using it, we provide the pricing formulae of the puttable and callable bonds with credit risk.\textbf{}



%
%

\section{Introduction}

\indent

Issuing bonds is one of financing tools of governments or firms and the corporate bonds are defaultable. The pricing problem for defaultable bonds that is related to credit risk modeling and the credit risk problem is one of the most interesting problems in financial mathematics.

Among the corporate bonds, there are some bonds with provisions related to early redemption \cite{Hul2006,BM2003}. There are two kinds of early redemption provisions. One is the provision allowing the bond holder to demand early redemption. In this case, the bond holder has the right to demand early redemption from the bond issuer prior to maturity. That is, the bond holder has the right to put (sell) the bond back to the bond issuer at a predetermined price (early redemption premium). Such a bond is called a \textit{puttable bond}. It can be seen that the holder of such a bond has purchased a bond put option which provides the right to sell the bond prior to maturity as well as the bond itself. Because the put option increases the value of the bond to the holder, the prices of the puttable bonds are more expensive than the straight bonds with no early redemption (put) features. Another one is the provision allowing the bond issuing firm to demand early ending the debt contract. In this case, the issuing firm has the right to buy back the bond at a predetermined price prior to maturity. Such a bond is called a \textit{callable bond}. It can be seen that the holder of such a bond has sold a call option on the bond to the bond issuer. The call price in this option is the predetermined price that must be paid by the issuer to the holder and thus the prices of the callable bonds are cheaper than bonds with no early redemption (call) features. 

It seems that it is clear why the bond holder demand early redemption. But what is the reason that the bond issuer demands such a call provision? One reason is that bond indentures often place a number of restrictions to the issuing firm and companies need some routes to escape from the restrictions if the restrictions prove too inhibiting. The call provision provides the escape route \cite{BM2003}. 

Puttable or callable bonds include some bond options as described in the above. Recently, pricing problems of bond options are widely studied. In \cite{FNV2016, ZEW2017, ZW2012} are studied the pricing problems of American bond options and in \cite{RF2016, RF2017} are considered the pricing problems of coupon bonds as interest derivatives under the 1 and 2 factor Hull-White models. These are all the pricing problems for bond options without credit risk.

According to experts, the pricing problem of puttable bonds(options on bonds) with credit risk is a challenging problem. \cite{OK2020} provides a partial differential equation model of the price of a Discrete Coupon Bond with Early redemption provision (with credit risk) and an analytical solution representation but their approach is to analyze together by adding an early-redemption provision to the pricing model of the discrete-coupon bonds of \cite{OJKJ2017, OKP2014}. That is not the approach from the viewpoint of the bond option. On the other hand, In \cite{OKC2021}, authors proposed a 1 factor structural model for option on zero coupon bond with credit risk and provided analytical pricing formulae of the puttable and callable bonds with credit risk using the solution representations and monotonicity result of the solutions to the boundary value problems for Black-Scholes equations with some special maturity payoff functions.

The purpose of this paper is to extend the 1 factor models and pricing formulae for options on bonds with credit risk provided in \cite{OKC2021} to the 2 factor models. To this end, first, a 2 factor structural (PDE) model of pricing zero coupon bond with credit risk is provided, the analytical pricing formula is derived under some conditions for default boundary and default recovery, and the strict monotonicity of the bond price function with respect to the firm value variable is proved. Then 2 factor structural PDE model of the option on zero coupon bond with credit risk is provided. Unlike the 1 factor model in \cite{OKC2021}, this model is a terminal boundary value problem for a partial differential equation with variable coefficients and two spatial variables and thus generally does not have analytical pricing formula. In this paper under some conditions on the default barrier and early redemption premium using zero coupon bond as numeraire, the 2 factor model is transformed into a terminal boundary value problem for Black-Scholes equation with time dependent coefficient and its analytical pricing formula is derived using the methods of \cite{Jia2005} and \cite{OKC2021}. Using it, we provide the pricing formulae of the puttable and callable bonds with credit risk. Due to the restrictions on the default barrier and early redemption premium, our pricing formulae do not include the results of \cite{OKC2021} but in the case that $q=0,\; \; a=r,\; \; E=Ee^{-r(T-T_{1} )} $ they include the results of \cite{OKC2021}.

Pricing problems of puttable or callable bonds are divided by pricing the straight bonds (the bonds without put or call provisions) and calculation of bond option premium. So the rest of the paper is organized as follows: in Section 2, a 2 factor structural PDE model for the straight bond with credit risk and the pricing formula are provided and the monotonicity of the bond price with respect to the firm value variable is proved. In Section 3, the pricing model for option on bond with credit risk and the pricing formulae of bond option and puttable (callable) bond with credit risk are provided. 

\section{Two Factor Structural Model for Zero Coupon Straight Bond with Credit Risk and Pricing Formula}

\subsection {Two factor structural model for zero coupon straight bond}

\textbf{Assumption 1. }Risk free rate follows \textit{Vasicek} model:
\begin{equation} \label{2.1} \tag{2.1}
dr_{t} =\theta (\mu -r_{t} )dt+s_{r} dW_{1} (t).                          
\end{equation} 
Here $W_{1} $ is standard Brownian motion and $\theta ,\; \; \mu ,\; \; s_{r} $ are constants. Under the condition \eqref{2.1}, the price $Z(r,\; \; t)=Z(r,\; \; t:\; T)$ of default free zero coupon bond with maturity $T$ is the solution to the problem
\begin{equation*} \label{2.2}\tag{2.2} 
\left\{\begin{array}{l} {\displaystyle\frac{\partial Z}{\partial t}+\frac{1}{2} s_{r} {}^{2} \displaystyle\frac{\partial ^{2} Z}{\partial {\kern 1pt} r^{2} } +\theta (\mu -r)\displaystyle\frac{\partial Z}{\partial {\kern 1pt} r} -rZ=0,} \\ {Z(r,\; \; T)=1} \end{array}\right.  
\end{equation*} 
and provided as follows (see \cite{Hul2006}.):
\begin{equation} \label{2.3} \tag{2.3}
Z(r,\; \; t)=Z(r,\; \; t:T)=e^{\bar{A}(t,\; \; T)-\bar{B}(t,\; \; T)r} , 
\end{equation} 
\begin{equation} \label{2.4}\tag{2.4} 
\begin{array}{l} {\bar{B}(t,\; T)=\displaystyle\frac{1-e^{-\theta (T-t)} }{\theta } ,} \\ {\bar{A}(t,\; T)=-\displaystyle\int _{t}^{T}[\theta \mu \bar{B}(s,\; T)-\displaystyle\frac{1}{2} s_{r} {}^{2} \bar{B}^{2} (s,\; T)]ds } \\ {\; \; \; \; \; \; =[\bar{B}(t,\; T)-T+t](\mu -\displaystyle\frac{s_{r} {}^{2} }{2\theta ^{2} } )-\displaystyle\frac{s_{r} {}^{2} }{4\theta } \bar{B}^{2} (t,\; T).} \end{array} 
\end{equation} 

\textbf{Assumption 2. }The firm value $V$consists of $m$ shares of traded asset $S$ and $n$ sheets of zero coupon bonds:
\begin{equation} \label{2.5}\tag{2.5} 
V_{t} =mS_{t} +nC_{t}  
\end{equation} 
and follows the geometric Brownian motion:
\begin{equation} \label{2.6} \tag{2.6}
\displaystyle\frac{dV(t)}{V(t)} =\beta (r_{t} ,V_{t} ,t)dt+s_{V} (r_{t} ,V_{t} ,t)dW_{2} (t).                 
\end{equation} 
Here $W_{2} $ is standard Brownian motion and the firm does not pay out any dividend.

\textbf{Assumption 3. }Default event occurs when $V\le V_{b} (r,\; \; t)$.

\textbf{Assumption 4. }Default recovery is $R_{d} (V_{t} ,\; \; r_{t} ,\; \; t)$.

\textbf{Assumption 5.}
\begin{equation}\label{2.7}\tag{2.7}
dW_{1} \cdot dW_{2} =\rho {\kern 1pt} dt  (ignoring\; higher\; orders\; of\; dt)
\end{equation}

\textbf{Assumption 6.} The price $C_{t} $ at time $t$ of our bond is provided by a deterministic function \textbf{$f(x_{1} ,x_{2} ,t)$ }as $C_{t} =f(r_{t} ,\; \; V_{t} ,\; \; t)$ and the price at time $T$ (the face value) is 1(unit of currency). For convenience, \textbf{$f(x_{1} ,x_{2} ,t)$ }is written as $C(r,\; \; V,\; \; t)$.

Under the assumptions 1$\mathrm{\sim}$6, the price function $C(r,\; \; V,\; \; t)$ satisfies the following PDE. (Note that this is a special case of the model studied in Section 5 of \cite{OW2013}.)
\begin{equation} \label{2.8}\tag{2.8} 
\displaystyle\frac{\partial C}{\partial t} +\displaystyle\frac{1}{2} \left(s_{r}^{2} \displaystyle\frac{\partial ^{2} C}{\partial r^{2} } +2\rho {\kern 1pt} s_{r} s_{V} V\displaystyle\frac{\partial ^{2} C}{\partial r\partial V} +s_{V}^{2} V^{2} \displaystyle\frac{\partial ^{2} C}{\partial V^{2} } \right)+\theta (\mu -r)\displaystyle\frac{\partial C}{\partial r} +rV\displaystyle\frac{\partial C}{\partial V} -rC=0.   
\end{equation} 
Thus the PDE model for our bond is the terminal value problem given in domain $\Sigma =$ $\{ (r,\; \; V,\; \; t):0<t<T,\; \; r>0,\; \; V>V_{b} (r,\; \; t)\} $ and consisting of \eqref{2.8} and the following terminal value condition and boundary condition:
\begin{equation} \label{2.9} \tag{2.9}
C(r,\; \; V,\; \; T)=1 
\end{equation} 
\[C(r,\; \; V_{b} (r,t),\; \; t)=R_{d} (V_{b} (r,t),\; \; r,\; \; t).\] 
This problem is terminal boundary value problem for a partial differential equation with variable coefficients and two spatial variables and thus generally does not have analytical pricing formula.

\subsection{Analytic pricing formula and monotonicity with respect to firm value variable}

To obtain the solution representation, some additional assumptions are added

\textbf{Assumption 7. }$V_{b} (r,\; t)={\rm \; }BZ(r,\; t)(B$ is a constant and $s_{V} $ is a constant, too. The default recovery is provided as follows (exogenous face value recovery)

\begin{equation}
R_{d} =R\cdot Z(r,\; \; t)(0\le R<1)(exogenous\; face\; value\; recovery).       \label{2.10}\tag{2.10}
\end{equation}

\noindent Under these assumptions, the domain and boundary condition are modified as follows:
\[\Sigma =\{ (r,\; \; V,\; \; t):0<t<T,\; \; r>0,\; \; V>BZ(r,\; \; t)\} \] 
\begin{equation} \label{2.11} \tag{2.11}
C(r,\; \; B\cdot Z(r,\; \; t),\; \; t)=R\cdot Z(r,\; \; t) 
\end{equation} 

Now find the solution representation of the terminal boundary value problem \eqref{2.8}, \eqref{2.9}, \eqref{2.11} in $\Sigma =\{ (r,\; \; V,\; \; t):0<t<T,\; \; r>0,\; \; V>BZ(r,\; \; t)\} $.

In the problem \eqref{2.8}, \eqref{2.9}, \eqref{2.11}, using the transformation 
\begin{equation} \label{2.12}\tag{2.12} 
x=\displaystyle\frac{V}{Z} ,\; \; u(x,\; \; t)=\displaystyle\frac{C(r,\; \; V,\; \; t)}{Z}  
\end{equation} 
(here zero coupon bond price $Z$ is used as numeraire) and considering \eqref{2.2} and $Z_{r} /Z=-\bar{B}(t,\; \; T)$, then we have the following problem
\[\begin{array}{l} {u_{t} +\displaystyle\frac{1}{2} \sigma _{x}^{2} (t)x^{2} u_{xx} =0,\; \; (x>B,\quad 0<t<T)} \\ {u(x,\; \; T)=1,\quad \quad \quad \quad \quad \quad \quad \quad \quad (x>B)} \\ {u(B,\; \; t)=R,\quad \quad \quad \quad \quad \quad \quad \quad (0<t<T).} \end{array}\] 
Here $\sigma _{x}^{} (t)$ is defined as follows:
\begin{equation} \label{2.13}\tag{2.13} 
\sigma _{x}^{2} (t\; ;\; T)=s_{r}^{2} \bar{B}^{2} (t,\; T)+s_{V}^{2} -2\rho {\kern 1pt} s_{r} \bar{B}(t,\; T)s_{V} \ge 0,\; \; 0\le t\le T. 
\end{equation} 
Using the transformation 
\begin{equation} \label{2.14}\tag{2.14} 
u=R+(1-R)W,                              
\end{equation} 
then we have the following problem for W.
\begin{equation} \label{2.15}\tag{2.15}
\begin{array}{l} {W_{t} +\displaystyle\frac{1}{2} \sigma _{x}^{2} (t)x^{2} W_{xx} =0,\; \; (x>B,\quad 0<t<T)} \\ {W(x,\; \; T)=1,\quad \quad \quad \quad \quad \quad \quad (x>B)} \\ {W(B,\; \; t)=0,\quad \quad \quad \quad \quad \quad \; \; (0<t<T).} \end{array} 
\end{equation} 
Returning to original variables by \eqref{2.12} and \eqref{2.14}, then our bond price can be represented by $W$ as follows:
\begin{equation} \label{2.16}\tag{2.16} 
C(r,\; \; V,\; \; t)=[W(1-R)+R]Z=W(V/Z,\; \; t)Z+[1-W(V/Z,\; \; t)]RZ.        
\end{equation} 

Now we solve \eqref{2.15}. Use the transformation
\[s=\displaystyle\int _{0}^{t}\sigma _{x}^{2} (u,T)du,\; \; T^{*} =\displaystyle\int _{0}^{T}\sigma _{x}^{2} (u,\; T)du ,\; \; W(x,\; \; t)=f(x,\; \; s) .\] 
Then \eqref{2.15} is changed to the following \cite{Jia2005}.
\[\begin{array}{l} {f_{s} +\displaystyle\frac{1}{2} x^{2} f_{xx} =0,\quad \quad (0<s<T^{*} ,\quad x>B)} \\ {f(B,\; \; s)=0,\quad \quad \quad \quad \quad \quad (0<s<T^{*} )} \\ {f(x,\; \; T^{*} )=1.\quad \quad \quad \quad \quad \quad (x>B).} \end{array}\] 
This problem is a special case of the problem (6), (7), (8) studied in \cite{OKC2021} and the solution is given by
\[f(x,s)=N(d_{1} )-\displaystyle\frac{x}{B} N(d_{2} ),\] 
and from Theorem 2 of \cite{OKC2021}, we have  
\[f_{x} (x,\; \; s)>0,\; \; 0=f(B,\; \; s)<f(x,\; \; s)<f(+\infty ,\; \; s)=1\; \; (x>B,\; \; s<T*).\] 
Here
\[N(x)=\displaystyle\frac{1}{\sqrt{2\pi } } \displaystyle\int _{-\infty }^{x}e^{-t^{2} /2} dt ,  d_{1} =\displaystyle\frac{\ln \displaystyle\frac{x}{B} -\displaystyle\frac{1}{2} (T^{*} -s)}{\sqrt{T^{*} -s} } ,\; \; d_{2} =\displaystyle\frac{\ln \displaystyle\frac{B}{x} -\displaystyle\frac{1}{2} (T^{*} -s)}{\sqrt{T^{*} -s} } .\] 
Returning to original variables, the we have the following representation and estimates for $W$.
\begin{equation} \label{2.17}\tag{2.17} 
\begin{array}{l} {W(x,\; \; t)=N(d_{1} )-\displaystyle\frac{x}{B} N(d_{2} ),} \\ {W_{x} (x,\; t)>0,\; \; 0=W(B,\; t)<W(x,\; t)<W(+\infty ,\; t)=1\; \; (x>B,\; 0\le t<T)} \end{array} 
\end{equation} 
\begin{equation} \label{2.18} \tag{2.18}
\begin{array}{l} {d_{1} =d(x/B,\; \; t,\; \; T):=\displaystyle\frac{\ln \displaystyle\frac{x}{B} -\displaystyle\frac{1}{2} \displaystyle\int _{t}^{T}\sigma _{x}^{2} (u,\; \; T)du }{\sqrt{\displaystyle\int _{t}^{T}\sigma _{x}^{2} (u,\; \; T)du } } =\displaystyle\frac{\ln \displaystyle\frac{V}{BZ} -\displaystyle\frac{1}{2} \displaystyle\int _{t}^{T}\sigma _{x}^{2} (u,\; \; T)du }{\sqrt{\displaystyle\int _{t}^{T}\sigma _{x}^{2} (u,\; \; T)du } } :=a,} \\ {d_{2} =d(B/x,\; \; t,\; \; T)=\displaystyle\frac{\ln \displaystyle\frac{B}{x} -\displaystyle\frac{1}{2} \displaystyle\int _{t}^{T}\sigma _{x}^{2} (u,\; \; T)du }{\sqrt{\displaystyle\int _{t}^{T}\sigma _{x}^{2} (u,\; \; T)du } } =\displaystyle\frac{\ln \displaystyle\frac{BZ}{V} -\displaystyle\frac{1}{2} \displaystyle\int _{t}^{T}\sigma _{x}^{2} (u,\; \; T)du }{\sqrt{\displaystyle\int _{t}^{T}\sigma _{x}^{2} (u,\; \; T)du } } :=\tilde{a}.} \end{array} 
\end{equation} 
Here zero coupon bond price $Z$ is given by \eqref{2.3}.

\textbf{Remark 1}. From \eqref{2.16}, the price of our bond at time $t$ can be seen as an expectation of the current value of the bond in the case that we have no default at or after time $t$ and the value of the bond in the case that default event occurs at or after time $t$. So $W(V/Z,\; \; t)$ in \eqref{2.16} can be regarded as the survival probability (probability of no default at or after time $t$) in the case with default barrier $BZ(r,\; \; t)$ and \eqref{2.15} is the PDE model for the survival probability at or after time $t$ in the case with default barrier $BZ(r,\; \; t)$.

Consider $x=V/Z$ and substitute \eqref{2.17} into \eqref{2.16}, then we have the following theorem.

\textbf{Theorem 1.} \textit{Under the assumptions 1$\sim$7 the solution to the problem \eqref{2.8}, \eqref{2.9}, \eqref{2.11} (the price function of zero coupon straight bond with default boundary $BZ(r,\; \; t)$) is represented as follows:}
\begin{equation} \label{2.19}\tag{2.19} 
C(r,\; \; V,\; \; t)=[R+W(V/Z,\; \; t)(1-R)]\cdot Z(r,\; \; t).                   
\end{equation} 
\textit{In particular,}$C(r,\; V,\; t)/Z(r,\; t)$ \textit{is an strictly increasing function of the variables }$x=V/Z$\textit{and furthermore we have }
\begin{equation} \label{2.20}\tag{2.20} 
C_{V} (r,\; \; V,\; \; t)>0,\; \; R\cdot Z(r,\; \; t)<C(r,\; \; V,\; \; t)<Z(r,\; \; t)\; \; (V>B\cdot Z(r,\; \; t),\; \; 0\le t<T).    
\end{equation} 
\textit{Thus the bond price \eqref{2.19} is strictly increasing on} $V(>BZ(r,\; \; t))$.

\textbf{Remark 2}. The financial meaning of \eqref{2.19} is clear: the first term is the current price of the part to be given to bond holder regardless of default event and the second term is the allowance dependent on default probability at time $t$. If at some moment $t$, the default is certain ($W(V/Z,\; t)=0$) = 0), then the price of the bond at $t$ is exactly the current price of default recovery. And as shown in formula \eqref{2.19}, if $R=0$, that is, if there is nothing to recover in case of default, then the bond price is the product of zero coupon bond and survival probability; and if $R=1$, that is, if there is no any loss in case of default, then the bond price is the same with risk free zero coupon bond.

\textbf{Remark 3}. Theorem 1 does not include Theorem 3 (for 1 factor model) of \cite{OKC2021} due to the peculiarity of default boundary (the assumption 7) but in the case with $a=r$ and $q=0$ it includes Theorem 3 of \cite{OKC2021}.

\section{Two Factor Structural Model for Option on Bond with Credit Risk and Pricing Formulae}

\subsection{Puttable Bond with Credit Risk}

Now we assume that the bond issuing company added to the bond defined by the Assumptions 1$\mathrm{\sim}$6 a new provision that allows the bond holder to demand early redemption at a predetermined date prior to the maturity.

\textbf{Assumption 8.} At a predetermined date , the bond holder can receive $K$(unit of currency) and end the bond contract.

By this assumption, the bond holder has the credit bond defined by assumptions 1$\mathrm{\sim}$6 together with the right to sell this bond for the exercise price $K$ at time $T_{1} $. Thus the Assumption 8 provides the bond holder the bond (put) option with maturity $T_{1} $. This is a bond option on zero coupon bond with credit risk. If the default event occurs, then the bond holder receives the default recovery and the bond contract is ended, and thus the bond option contract is also ended. Therefore this bond option contract is a kind of barrier option contract and its price $P(r,\; \; V,\; \; t)$ is the solution to the following problem in domain $\Sigma _{1} =\{ (r,\; \; V,$$t):0<t<T_{1} ,\; \; r>0,\; \; V>V_{b} (r,\; \; t)\} $
\begin{equation} \label{3.1}\tag{3.1} 
\displaystyle\frac{\partial P}{\partial t} +\displaystyle\frac{1}{2} \left(s_{r}^{2} \displaystyle\frac{\partial ^{2} P}{\partial r^{2} } +2\rho {\kern 1pt} s_{r} s_{V} V\displaystyle\frac{\partial ^{2} P}{\partial r\partial V} +s_{V}^{2} V^{2} \displaystyle\frac{\partial ^{2} P}{\partial V^{2} } \right)+\theta (\mu -r)\displaystyle\frac{\partial P}{\partial r} +rV\displaystyle\frac{\partial P}{\partial V} -rP=0 
\end{equation} 
\[P(r,\; \; V_{b} (r,\; \; t),\; \; t)=0,\; \; \; \; \; \; \; \; \; \; \; \; 0\le t<T_{1} \] 
\[P(r,\; \; V,\; T_{1} )=(K-C(t,\; \; V,\; T_{1} ))^{+} ,\; \; V>V_{b} (T_{1} )\] 
This problem is a boundary value problem of parabolic equation with 2 spatial variables and moving boundary and thus generally, it does not have solution representation.

In order to obtain its solution representation, we add the assumption 7 in Section 2 and the following assumption.

\textbf{Assumption 9.} $K=E\cdot Z(r,\; T_{1} )\; (E$: constant)

Then the domain of the problem is $\Sigma _{1} =\{ (r,\; \; V,\; \; t):0<t<T_{1} ,\; \; r>0,\; \; V>BZ(r,\; \; t)\} $ and the boundary conditions are changed as follows:
\begin{equation} \label{3.2}\tag{3.2} 
P(r,\; \; BZ(r,\; \; t),\; \; t)=0,\; \; \; \; \; \; \; \; \; \; \; \; 0\le t<T_{1}  
\end{equation} 
\[P(r,\; \; V,\; \; T_{1} )=(EZ(r,\; \; T_{1} )-C(r,\; \; V,\; T_{1} ))^{+} ,\; \; V>BZ(r,\; \; T_{1} ).\] 
Here$R<C(r,\; V,\; T_{1} )/Z(r,\; T_{1} )<1$ from \eqref{2.19} and \eqref{2.17} and from Theorem 1, 
\[C(r,\; V,\; T_{1} )/Z(r,\; T_{1} )=R+W(V/Z,\; \; t)(1-R)\] 
is a strict increasing function with respect to $V/Z$. Thus if$R<E<1$, then there exists a constant $L>B$ such that $R+W(L,\; \; t)(1-R)=E$ and 

$C(r,\; \; V,\; T_{1} )<E\cdot Z(r,\; T_{1} )$ for $V<L\cdot Z(r,\; T_{1} )$,  

$C(r,\; \; V,\; T_{1} )\ge E\cdot Z(r,\; T_{1} )$ for $V\ge L\cdot Z(r,\; T_{1} )$. 

\noindent Thus the terminal condition of the bond option is written as follows:
\[P(r,\; \; V,\; \; T_{1} )=[EZ(r,\; \; T_{1} )-C(r,\; \; V,\; T_{1} )]\cdot 1\{ V<L\cdot Z(r,\; \; T_{1} )\} ,\; \; V>BZ(r,\; \; T_{1} ).\] 
($L\cdot Z(r,\; T_{1} )$ is called the early redemption boundary at time $T_{1} $.) Therefore considering \eqref{2.16}, then we have
\begin{equation} \label{3.3}\tag{3.3} 
P(r,\; \; V,\; \; T_{1} )=Z(r,\; \; T_{1} )\left\{E-R-W(\displaystyle\frac{V}{Z} ,\; \; t)(1-R)]\right\}\cdot 1\left\{\displaystyle\frac{V}{Z(r,\; \; T_{1} )} <L\right\},\; \; \displaystyle\frac{V}{Z(r,\; \; T_{1} )} >B.   
\end{equation} 
Now we find the solution representation of the problem \eqref{3.1}, \eqref{3.2}, \eqref{3.3}. Use the transformation $x=V/Z,\; \; p(x,\; \; t)=P(r,\; \; V,\; \; t)/Z$, then we have
\begin{equation} \label{3.4} \tag{3.4}
p_{t} +\displaystyle\frac{1}{2} \sigma _{x}^{2} (t)x^{2} p_{xx} =0,\; \; \; \; \; \; \; \; \; \; \; \; \; \; \; \; \; \; \; \; \; \; \; \; \; \; \; \; \; \quad (x>B,\; \; 0<t<T_{1} ) 
\end{equation} 
\begin{equation} \label{3.5}\tag{3.5}
p(B,\; \; t)=0,\quad \quad \quad \quad \quad \quad \quad \qquad  \; \; \; \; \; \; \; \; \; \; \; \; \; \; \; \; \; \; \; \; \; \; \; \; \quad (0<t<T_{1} ) 
\end{equation} 
\[p(x,\; \; T_{1} )=[E-R-(1-R)W(x,\; \; T_{1} )]\cdot 1\{ x<L\} ,\; \; \; \; (x>B).\] 
Consider \eqref{2.17}, then we have
\begin{equation} \label{3.6}\tag{3.6} 
p(x,\; T_{1} )=\left\{E-R-(1-R)\left[N(\bar{d}_{1} )-\left(\displaystyle\frac{x}{B} \right)N(\bar{d}_{2} )\right]\right\}\cdot 1\{ x<L\} ,\; \; (x>B).   
\end{equation} 
Here
\begin{equation} \label{3.7}\tag{3.7} 
\bar{d}_{1} =d(x/B,T_{1} ,T)=\left(\sqrt{\displaystyle\int _{T_{1} }^{T}\sigma _{x}^{2} (u,T)du } \right)^{-1} \left(\ln \displaystyle\frac{x}{B} -\displaystyle\frac{1}{2} \displaystyle\int _{T_{1} }^{T}\sigma _{x}^{2} (u,T)du \right),\bar{d}_{2} =d(B/x,T_{1} ,T). 
\end{equation} 
As in Section 2, use the transformation
\[s=\displaystyle\int _{0}^{t}\sigma _{x}^{2} (u,T)du,\; \; p(x,\; \; t)=f(x,\; \; s),\; \; T_{1}^{*} =\displaystyle\int _{0}^{T_{1} }\sigma _{x}^{2} (u,\; T)du  ,\] 
then we have
\begin{equation} \label{3.8}\tag{3.8} 
f_{s} +\displaystyle\frac{1}{2} x^{2} f_{xx} =0,\; \; 0<s<T_{1}^{*} ,\; \; x>B 
\end{equation} 
\begin{equation} \label{3.9}\tag{3.9} 
\begin{array}{l} {f(x,\; \; T_{1}^{*} )=[E-R]\cdot 1\{ B<x<L\} -(1-R)N(\bar{d}_{1} )\cdot 1\{ B<x<L\} } \\ {\; \; \; \; \; \; \; \; \; \; \; \; \; \; +(1-R)\left(\displaystyle\frac{x}{B} \right)N(\bar{d}_{2} )\cdot 1\{ B<x<L\} ,\; \; \; \; x>B} \end{array} 
\end{equation} 
\begin{equation} \label{3.10}\tag{3.10} 
f(B,\; \; s)=0,\; \; 0<s<T_{1}^{*} .                                   
\end{equation} 
Here
\[\begin{array}{l} {\bar{d}_{1} (x\; ;\; \; T_{1}^{*} ,\; \; T^{*} )=\displaystyle\frac{\ln \displaystyle\frac{x}{B} -\displaystyle\frac{1}{2} (T^{*} -T_{1}^{*} )}{\sqrt{T^{*} -T_{1}^{*} } } (=d(x/B\; ;\; T_{1} ,\; \; T))\; ,} \\ {\bar{d}_{2} (x\; ;\; \; T_{1}^{*} ,\; \; T^{*} )=\displaystyle\frac{\ln \displaystyle\frac{B}{x} -\displaystyle\frac{1}{2} (T^{*} -T_{1}^{*} )}{\sqrt{T^{*} -T_{1}^{*} } } (=d(B/x\; ;\; T_{1} ,\; \; T))\; .} \end{array}\] 
The problem \eqref{3.8}$\mathrm{\sim}$\eqref{3.10} is a special case of the problem (19)$\mathrm{\sim}$(21) of \cite{OKC2021}(with $r=q=a=0,\; \; \sigma =1,\; \; K=L$, $T_{1} =T_{1}^{*} ,\; \; T=T^{*} $) and thus from Theorem 4 of \cite{OKC2021}, the solution to the problem \eqref{3.8}$\mathrm{\sim}$\eqref{3.10} is represented as follows:
 
\[\begin{array}{l} {f(x,\; \; s)=(E-R)\{ N(b_{1} (x))-N(b_{2} (x))-(x/B)[N(\tilde{b}_{1} (x))-N(\tilde{b}_{2} (x))\} } \\ {\; \; \; \; \; \; \; \; \; \; -(1-R)\{ N_{2} (a(x),\; \; b_{1} (x);\; \; \delta )-N_{2} (a(x),\; \; b_{2} (x);\; \; \delta )-} \\ {\; \; \; \; \; \; \; \; \; \; -(x/B)[N_{2} (\tilde{a}(x),\; \; \tilde{b}_{1} (x);\; \; \delta )-N_{2} (\tilde{a}(x),\; \; \tilde{b}_{2} (x);\; \; \delta )]\} } \\ {\; \; \; \; \; \; \; \; \; \; +(1-R)\{ (x/B)[N_{2} (\tilde{a}(x),\; \; -\tilde{b}_{1} (x);\; \; -\delta )-N_{2} (\tilde{a}(x),\; \; -b_{3} (x);\; \; -\delta )]-} \\ {\; \; \; \; \; \; \; \; \; \; \; \; \; \; -[N_{2} (a(x),\; \; -b_{1} (x);\; \; -\delta )-N_{2} (a(x),\; \; -b_{3} (x);\; \; -\delta )]\} ,\; \; x>B,\; \; s<T_{1}^{*} .} \end{array}\] 
Here
\[b_{1} (x)=\displaystyle\frac{\ln \displaystyle\frac{x}{B} -\displaystyle\frac{1}{2} (T_{1}^{*} -s)}{\sqrt{T_{1}^{*} -s} } =d(x/B,\; \; t,\; \; T_{1} ),\; \; \; \; b_{2} (x)=\displaystyle\frac{\ln \displaystyle\frac{x}{L} -\displaystyle\frac{1}{2} (T_{1}^{*} -s)}{\sqrt{T_{1}^{*} -s} } =d(x/L,\; \; t,\; \; T_{1} ),\] 
\[\tilde{b}_{1} (x)=b_{1} (B^{2} /x)=d(B/x,\; \; t,\; \; T_{1} ),\; \; \; \; \; \; \; \; \; \; \; \; \; \; \; \tilde{b}_{2} (x)=b_{2} (B^{2} /x)=d(B^{2} /Lx,\; \; t,\; \; T_{1} ),\] 
\[a(x)=\bar{d}_{1} (x\; ;\; \; s,\; \; T^{*} )=d(x/B,\; \; t,\; \; T)\; ,\; \; \; \; \; \; \; \; \; \; \; \tilde{a}(x)=\bar{d}_{2} (x\; ;\; \; s,\; \; T^{*} )=d(B/x,\; \; t,\; \; T)\; ,\] 
\[\tilde{b}_{3} (x)=-\displaystyle\frac{\ln \displaystyle\frac{x}{L} +\displaystyle\frac{1}{2} (T_{1}^{*} -s)}{\sqrt{T_{1}^{*} -s} } =d(L/x,\; \; t,\; \; T_{1} ),\; \; b_{3} (x)=\tilde{b}_{3} (B^{2} /x)=d(Lx/B^{2} ,\; \; t,\; \; T_{1} ),\] 
\begin{equation} \label{3.11}\tag{3.11} 
\delta (s)=\sqrt{(T_{1}^{*} -s)/(T^{*} -s)} =\sqrt{{\displaystyle\int _{t}^{T_{1} }\sigma _{x}^{2} (u,\; T)du  \mathord{\left/{\vphantom{\displaystyle\int _{t}^{T_{1} }\sigma _{x}^{2} (u,\; T)du  \displaystyle\int _{t}^{T}\sigma _{x}^{2} (u,\; T)du }}\right.\kern-\nulldelimiterspace} \displaystyle\int _{t}^{T}\sigma _{x}^{2} (u,\; T)du } } =\bar{\delta }(t).                
\end{equation} 
Returning to the original variables, we have the following solution representation of the problem \eqref{3.1}, \eqref{3.2}, \eqref{3.3} in the domain $\Sigma _{1} =\{ (r,\; \; V,\; \; t):0<t<T_{1} ,\; \; r>0,\; \; V>BZ(r,\; \; t)\} $:
\[P(r,\; \; V,\; t)=p(V/Z,\; \; t)\cdot Z(r,\; \;t )=\] 
\[\begin{array}{l} {=Z(r,\; \;t )(E-R)\left\{N(b_{1} )-N(b_{2} )-\displaystyle\frac{V}{BZ} [N(\tilde{b}_{1} )-N(\tilde{b}_{2} )\right\}} \\ {\; \; -Z(r,\; \;t )(1-R)\left\{N_{2} (a,\; \; b_{1} ;\; \; \bar{\delta })-N_{2} (a,\; \; b_{2} ;\; \; \bar{\delta })-\displaystyle\frac{V}{BZ} [N_{2} (\tilde{a},\; \; \tilde{b}_{1} ;\; \; \bar{\delta })\right. } \\ {\; \; \; \; \; \; \; \; \; \; \; \; \; \; \; \; \; \; \; \; \; \; \; \; \; \; \; -N_{2} (\tilde{a},\; \; \tilde{b}_{2} ;\; \; \bar{\delta })]+N_{2} (a,\; \; -b_{1} ;\; \; -\bar{\delta })-N_{2} (a,\; \; -b_{3} ;\; \; -\bar{\delta })} \\ {\; \; \; \; \; \; \; \; \; \; \; \; \; \; \; \; \; \; \; \; \; \; \; \; \; \; \; \left. -\displaystyle\frac{V}{BZ} [N_{2} (\tilde{a},\; \; -\tilde{b}_{1} ;\; \; -\bar{\delta })-N_{2} (\tilde{a},\; \; -\tilde{b}_{3} ;\; \; -\bar{\delta })]\right\}.} \end{array}\] 
\begin{equation} \label{3.12}\tag{3.12} 
\begin{array}{l} {=Z(r,\; \; t)\cdot \{ (E-R)[N(b_{1} )-N(b_{2} )]-(1-R)[N_{2} (a,\; \; b_{1} ;\; \; \bar{\delta })-N_{2} (a,\; \; b_{2} ;\; \; \bar{\delta })} \\ {\; \; \; \; \; \; \; \; \; \; \; \; \; \; \; \; \; \; \; \; \; \; \; \; \; \; \; \; \; \; \; \; \; \; \; \; \; \; \; \; \; \; \; +N_{2} (a,\; \; -b_{1} ;\; \; -\bar{\delta })-N_{2} (a,\; \; -b_{3} ;\; \; -\bar{\delta })]\} } \\ {\; \; +\displaystyle\frac{V}{B} \{ -(E-R)[N(\tilde{b}_{1} )-N(\tilde{b}_{2} )+(1-R)[N_{2} (\tilde{a},\; \; \tilde{b}_{1} ;\; \; \bar{\delta })-N_{2} (\tilde{a},\; \; \tilde{b}_{2} ;\; \; \bar{\delta })} \\ {\; \; \; \; \; \; \; \; \; \; \; \; \; \; \; \; \; \; \; \; \; \; \; \; \; \; \; \; \; \; \; \; \; \; \; \; \; \; \; +N_{2} (\tilde{a},\; \; -\tilde{b}_{1} ;\; \; -\bar{\delta })-N_{2} (\tilde{a},\; \; -\tilde{b}_{3} ;\; \; -\bar{\delta })]\} } \end{array} 
\end{equation} 
Here
\begin{equation} \label{3.13} \tag{3.13}
\begin{array}{l} {d(x,\; \; t,\; \; T):=\left(\sqrt{\displaystyle\int _{t}^{T}\sigma _{x}^{2} (u,\; \; T)du } \right)^{-1} \left[\ln x-\displaystyle\frac{1}{2} \displaystyle\int _{t}^{T}\sigma _{x}^{2} (u,\; \; T)du \right]\; ,} \\ {a=d(V/(BZ),\; \; t,\; \; T)\; ,\; \; \; \; \; \; \; \; \; \tilde{a}=d(BZ/V,\; \; t,\; \; T)\; ,} \\ {b_{1} =d(V/(BZ),\; \; t,\; \; T_{1} )\; ,\; \; \; \; \; \; \; b_{2} =d(V/(LZ),\; \; t,\; \; T_{1} ),} \\ {\tilde{b}_{1} =d(BZ/V,\; \; t,\; \; T_{1} ),\; \; \; \; \; \; \; \; \; \tilde{b}_{2} =d(B^{2} Z/(LV),\; \; t,\; \; T_{1} )\; ,} \\ {b_{3} =d(LV/(B^{2} Z),\; \; t,\; \; T_{1} ),\; \; \; \tilde{b}_{3} =d(LZ/V,\; \; t,\; \; T_{1} )\; .} \end{array} 
\end{equation} 
Therefore we have proved the following theorem.

\textbf{Theorem 2.} Under the assumptions 1$\mathrm{\sim}$9, if $R<E<1$, then the price for bond put option (the solution to the problem \eqref{3.1}, \eqref{3.2}, \eqref{3.3} is given by \eqref{3.12}.

\textbf{Remark 4.} Assumption 8 gives the holder of the bond with expiry date $T$ the right to be able to demand early repayment of debts at the date $T_{1} $ prior to the expiry date (the early redemption recovery is$E\cdot Z(r,\; \; T_{1} )$). \eqref{3.12} is just the \textit{early redemption premium} which the holder should pay more due to this early redemption provision. Thus the price of the \textit{puttable bond} (the \textit{bond with early redemption provision}) in the time interval $[0,\; \; T_{1} ]$ is the sum of $B$ given by \eqref{2.19} and $P$ given by \eqref{3.13}, and the price in the time interval $(T_{1} ,\; \; T]$ is equal to $B$.

\textbf{Remark 5}. Theorem 2 does not include Theorem 4 (for 1 factor model) of \cite{OKC2021} due to the peculiarity of default boundary and early redemption recovery(the assumption 7 and 9) but in the case with $a=r,\; \; q=0$ and $E:=Ee^{-r(T-T_{1} )} $, it includes Theorem 4 of \cite{OKC2021}.

\subsection{Callable Bond with Credit Risk}

Now, we assume that the bond issuing company added to the bond defined by the Assumptions 1$\mathrm{\sim}$6 a new provision that allows the company to pay back the debt early at a predetermined date prior to the maturity.

\textbf{Assumption 10}: At a predetermined date , the bond issuing company can give $K$(unit of currency) and call back the bond.

\noindent By this provision, the bond issuing company has the right to purchase this bond defined by the Assumptions 1$\mathrm{\sim}$6 for the exercise price $K$ at time $T_{1} $. Thus the Assumption 10 provides the company the bond (call) option with maturity $T_{1} $. This is a bond option on zero coupon bond with credit risk. If the default event occurs, then the bond holder receives the default recovery and the bond contract is ended, and thus the bond option contract is ended, too. Therefore this bond option contract is a kind of barrier option contract and its price function $P=P(r,\; \; V,\; \; t)(0\le t\le T_{1} )$ satisfies \eqref{3.1} in the domain $\Sigma _{1} =\{ (r,\; V,$$t):0<t<T_{1} ,\; r>0,\; V>V_{b} (r,\; t)\} $ and boundary conditions similar to those in the model for puttable bond. This problem is a boundary value problem of parabolic equation with 2 spatial variables and moving boundary and thus generally, it does not have solution representation. In order to obtain its solution representation, we add the assumption 7 in Section 2 and the assumption 9 in the above. Then $P=P(r,\; \; V,\; \; t)$ $(0\le t\le T_{1} )$ satisfies \eqref{3.1} and \eqref{3.2} in $\Sigma _{1} =\{ (r,\; \; V,\; \; t):0<t<T_{1} ,\; \; r>0,\; \; V>BZ(r,\; \; t)\} $ and the following terminal condition:
\[P(r,\; \; V,\; \; T_{1} )=(C(r,\; V,\; T_{1} )-EZ(r,\; T_{1} ))^{+} ,\; \; V>BZ(r,\; \; T_{1} ). \] 
Here $C(r,\; V,\; T_{1} )$ is provided by \eqref{2.19}. Thus as in the precious subsection, if $R<E<1$, then there exists a constant $L>B$ such that the terminal condition of this bond option is written as follows:
\[P(r,\; \; V,\; \; T_{1} )=[C(r,\; \; V,\; T_{1} )-EZ(r,\; \; T_{1} )]\cdot 1\{ V>L\cdot Z(r,\; \; T_{1} )\} ,\; \; V>BZ(r,\; \; T_{1} ).\] 
Therefore from \eqref{2.19} and \eqref{2.17}, we can rewrite as follows:

\begin{equation} \label{3.14}\tag{3.14} 
\begin{array}{l} {P(r,\; \; V,\; \; T_{1} )=[R+W(\displaystyle\frac{V}{Z(r,T_{1} )} ,\; T_{1} )(1-R)-E]\cdot Z(r,\; T_{1} )\cdot 1\{ \displaystyle\frac{V}{Z(r,\; T_{1} )} >L\} ,\; \; \displaystyle\frac{V}{Z(r,\; T_{1} )} >B} \\ {\; \; \; \; \; \; \; \; \; \; \; \; \; \; \; =Z(r,\; T_{1} )\cdot \left\{R-E+(1-R)\left[N(d_{1} )-\displaystyle\frac{V}{BZ(r,\; T_{1} )} N(d_{2} )\right]\right\}\cdot 1\{ \displaystyle\frac{V}{Z(r,\; T_{1} )} >L\} .} \end{array} 
\end{equation} 

Using the similar method in Theorem 2 we have the following theorem.

\textbf{Theorem 3.} Under the assumptions 1$\mathrm{\sim}$7 and 9$\mathrm{\sim}$10, if $R<E<1$, then the price function for bond call option (the solution to the problem \eqref{3.1}, \eqref{3.2}, \eqref{3.14}) is provided as follows:
\[\begin{array}{l} {P(r,\; \; V,\; t)=Z(r,\; t)\left\{(R-E)\left[N(b_{2} )-\displaystyle\frac{V}{BZ} N(\tilde{b}_{2} )\right]\right. +} \\ {\; \; \; \; \; \; \; \; +(1-R)\left. \left[N_{2} (a,\; \; b_{2} ;\; \bar{\delta })-\displaystyle\frac{V}{BZ} N_{2} (\tilde{a},\; \; \tilde{b}_{2} ;\; \bar{\delta })+N_{2} (a,\; \; -b_{3} ;\; -\bar{\delta })-\displaystyle\frac{V}{BZ} N_{2} (\tilde{a},\; \; -\tilde{b}_{3} ;\; -\bar{\delta })\right]\right\}} \end{array}\] 
\begin{equation} \label{3.15}\tag{3.15} 
\begin{array}{l} {=Z(r,\; t)\{ (R-E)N(b_{2} )+\; (1-R)[N_{2} (a,\; \; b_{2} ;\; \bar{\delta })+N_{2} (a,\; \; -b_{3} ;\; -\bar{\delta })]\} +} \\ {\; \; \; -\displaystyle\frac{V}{B} \{ (R-E)N(\tilde{b}_{2} )\} +(1-R)[N_{2} (\tilde{a},\; \; \tilde{b}_{2} ;\; \bar{\delta })+N_{2} (\tilde{a},\; \; -\tilde{b}_{3} ;\; -\bar{\delta })]\} .} \end{array} 
\end{equation} 

\textbf{Remark 6.} Assumption 10 gives the bond issuing company the right to be able to pay$K$(unit of currency) and call back the bond with expiry date $T$ and puts the bond holder under an obligation. \eqref{3.15} is just the \textit{early redemption premium} which the bond issuing company should pay the bond holder due to this early redemption provision. Thus the price of the \textit{callable bond} (the \textit{bond with early redemption provision}) in the time interval $[0,\; \; T_{1} ]$ is the difference of $P$ given by \eqref{3.13} from $B$ given by \eqref{2.19}, and the price in the time interval $(T_{1} ,\; \; T]$ is equal to $B$.

\textbf{Remark 7}. Theorem 3 does not include Theorem 5 (for 1 factor model) of \cite{OKC2021} due to the peculiarity of default boundary and early redemption recovery(the assumption 7 and 9) but in the case with $a=r,\; \; q=0$ and $E:=Ee^{-r(T-T_{1} )} $, it includes Theorem 5 of \cite{OKC2021}.\textbf{}

\end{document}